\documentclass[unnumsec,webpdf,contemporary,large]{preprint-authoring-template}

\usepackage{tabularx,booktabs}
\usepackage{url}
\usepackage{fullpage}
\usepackage{graphicx}
\usepackage{lipsum}
\usepackage{titlesec}
\usepackage{upgreek}
\usepackage{todonotes}
\usepackage{soul}
\usepackage{subfig}
\usepackage{threeparttable}

\newcommand{\junk}[1]{}
\newcommand{\uop}{$\upmu$Op}

\begin{document}

\journaltitle{Bioinformatics}
\DOI{DOI HERE}
\copyrightyear{2026}
\pubyear{2026}
\access{Advance Access Publication Date: Day Month Year}
\appnotes{Paper}

\firstpage{1}

\title[aaa]{In-depth data movement characterization of DNA read mapping algorithms}

\author[1, $\ast$]{Berkan Şahin\ORCID{0009-0002-5151-2973}}
\author[1]{Ecem İlgün\ORCID{0000-0001-6496-6224}}
\author[1]{Klea Zambaku\ORCID{0000-0001-7951-7720}}
\author[1 $\ast$]{Can Alkan\ORCID{0000-0002-5443-0706}}


\address[1]{\orgdiv{Department of Computer Engineering}, \orgname{Bilkent University}, \orgaddress{\street{Bilkent}, \postcode{06800}, \state{Ankara}, \country{Türkiye}}}

\corresp[$\ast$]{Corresponding authors. \href{email:berkan.sahin@bilkent.edu.tr}{berkan.sahin@bilkent.edu.tr}, \href{email:calkan@cs.bilkent.edu.tr}{calkan@cs.bilkent.edu.tr}}

\received{Date}{0}{Year}
\revised{Date}{0}{Year}
\accepted{Date}{0}{Year}

\abstract{Read mapping algorithms represent a critical bottleneck in genomic analysis pipelines due to the resources required; a problem that is becoming more pronounced as sequencing technologies advance and datasets expand. We present an extensive workload characterization of leading read mapping algorithms across diverse sequencing technologies to identify hardware acceleration and optimization opportunities. We analyzed the following implementations: MashMap, wfmash, and Minimap2 for whole-genome alignment (WGA); NGMLR, Minimap2, and LRA for long reads; and Bowtie2, BWA-MEM, StrobeAlign, and minibwa for short reads.
We used the Intel VTune Profiler for detailed memory analysis to identify memory-bound functions, cache usage patterns, and data transfer bottlenecks. 
 We further analyzed a selected subset of alignment tools using the DAMOV-SIM simulator to obtain a more granular view of data movement characteristics for two different cache structures. 
 Based on our results, Minimap2 emerges as the fastest alignment tool for long read scenarios, but falls behind in whole-genome alignment due to poor multithreading. Across all three categories, some programs (e.g., LRA, BWA-MEM) are highly memory-bound. More in-depth analysis via DAMOV-SIM revealed that some memory-bound tools, in particular, do not fully utilize the multi-level cache hierarchy in modern multi-core CPU architectures. Our results show that while some tools require algorithmic redesign for performance gains, some others can benefit from running on PIM architectures.}
 \keywords{benchmarking, read mapping, long reads, short reads, processing in memory, genomics}

\maketitle

\section{Introduction}
    High-throughput sequencing (HTS) technologies 
revolution\-ized genomics by offering large-scale, affordable, and accurate sequencing of entire genomes. 
Read mapping is the process of aligning sequencing data to a reference genome. 
However, the massive amount of data produced by modern sequencers poses a significant computational challenge for read mapping. 
Even after decades of progress, read mapping remains a bottleneck in many genomic pipelines due to its resource intensiveness and growing algorithmic complexity. 
A key reason read mapping remains a resource-intensive workload is the cost of moving data between memory and processor cores, which creates a data movement bottleneck. 
This problem is heightened by large search spaces and irregular memory access patterns inherent to genome alignment. 
Processing-in-memory (PIM) architectures emerged as a promising solution to the data movement bottleneck~\cite{mutlu_primer_2024}. 
PIM architectures can drastically reduce the latency costs associated with transferring data between memory and the CPU by integrating the two, placing computational logic directly within memory units~\cite{biopim_position_2026}.
While prior work such as~\cite{boroumand_google_2018} demonstrated the potential of PIM for accelerating data-intensive workloads, or even more specifically, read mappers~\cite{realpim_2023}, and
Appuswamy et. al.~\cite{appuswamy_sequence_2018} analyzed sequence aligners from a microarchitectural perspective; their analysis was limited to single-core performance. There still remains a lack of detailed workload characterization of read mappers that takes multi-core performance and data movement patterns into account. Such an analysis would highlight how current algorithms could be adapted to leverage this paradigm and aid the development of future algorithms that account for data movement.

In this study, we aim to bridge this gap by conducting a comprehensive workload characterization of widely used alignment tools. Our goal is to understand which components of current algorithms are most constrained by memory access and data movement and to assess them as candidates for future PIM acceleration. To this end, we profile alignment tools across three categories: whole genome aligners, long read aligners, and short read aligners.

Whole genome aligners are represented by \texttt{minimap2}~\cite{minimap2}, \texttt{mashmap}~\cite{mashmap_2017}, and \texttt{wfmash}~\cite{wfmash_2021, wfalib_2021, wfalib_2023}; long read aligners by \texttt{NGMLR}~\cite{NGMLR}, \texttt{minimap2}~\cite{minimap2}, and \texttt{LRA}~\cite{LRA}; and short read aligners by \texttt{BWA-MEM}~\cite{bwamem_2013}, \texttt{bowtie2}~\cite{bowtie_2012}, \texttt{StrobeAlign}~\cite{strobealign_2026} and \texttt{minibwa}~\cite{minibwa}. 

We follow a two-phase approach: we first perform performance profiling with Intel VTune, and if the tool indicates a moderate to high percentage of memory-bound components, we follow up with DAMOV-SIM~\cite{oliveira_damov_2021} for further analysis.

\section{Methods}
    \subsection{Analyzed tools}

We analyze the following DNA read mapping tools:
\begin{itemize}
    \item \textbf{Whole genome alignment (WGA):} \texttt{minimap2}~\cite{minimap2}, \texttt{mashmap}~\cite{mashmap_2017}, \texttt{wfmash}~\cite{wfmash_2021, wfalib_2021, wfalib_2023}
    \item \textbf{Long read alignment:} \texttt{NGMLR}~\cite{NGMLR}, \texttt{minimap2}~\cite{minimap2}, \texttt{LRA}~\cite{LRA}
    \item \textbf{Short read alignment:} \texttt{BWA-MEM}~\cite{bwamem_2013}, \texttt{bowtie2}~\cite{bowtie_2012}, \texttt{StrobeAlign}~\cite{strobealign_2026}, \texttt{minibwa}~\cite{minibwa}
\end{itemize}

See the Supplementary Material for the program versions and commands used to invoke the tools.

\subsection{Top-down analysis}
We use Intel VTune's memory access mode to characterize a given program's data movement patterns. VTune is a profiler utilizing performance counters on Intel CPUs to report a hardware-based performance characterization, using a technique called top-down performance analysis, first proposed by Yasin~\cite{yasin_top-down_2014}. 

This analysis method classifies the time spent on each pipeline slot of a superscalar x86 CPU as follows~\cite{yasin_top-down_2014}:
\begin{enumerate}
    \item \textbf{Frontend bound:} This category denotes any pipeline stalls while instructions are being fetched, parsed, or decoded into micro-operations (\uop s).
    \item \textbf{Bad speculation:} This category denotes \uop s that are flushed out of the pipeline due to branch misprediction.
    \item \textbf{Backend bound:} This category denotes \uop s that are not accepted to the ``back end'' of the pipeline due to resource constraints. The two constraints most relevant to our analysis are the following: 
    \begin{enumerate}
        \item \textbf{Memory bound:} The CPU does not process the \uop~on the current clock cycle since the operands have not arrived from the memory yet.
        \item \textbf{Core bound:} The CPU cannot process the \uop~since all the ALUs are currently busy processing other \uop s.
    \end{enumerate}
    \item \textbf{Retiring:} This category denotes \uop s that the CPU finished processing and are being ``retired''. A high percentage of retired operations indicates that the workload utilizes the CPU efficiently and achieves throughput close to the theoretical maximum for the given processor.
\end{enumerate}

Given our characterization's focus on data movement, we are most concerned with the memory-boundness percentage for the entire workload and for each function. We use these metrics to select memory-bound functions in the target tools for further analysis using DAMOV-SIM. We use the same threshold as the DAMOV~\cite{oliveira_damov_2021} benchmark suite: we consider functions that take at least 3\% of total CPU time and are at least 30\% memory-bound.

\subsection{Function-level in-depth analysis}
To understand the data-movement characteristics of specific, critical functions, we use the DAMOV-SIM simulator in the DAMOV benchmark suite. DAMOV-SIM combines the Zsim x86 simulator with Ramulator, a memory simulator capable of simulating a diverse array of memory and cache configurations. We chose to use DAMOV-SIM for the function-level analysis because simulating all memory accesses provides a cleaner view of the data movement patterns of a given function, without the influence of other processes running on the test system or of parts of the workload we are not interested in. Furthermore, DAMOV-SIM allows us to experiment with different cache hierarchy configurations to see whether they alleviate data movement bottlenecks, as well as to simulate configurations with more CPU cores than the host system, thanks to Zsim.

\begin{figure}[hpt]
    \centering
    \subfloat[Host configuration]{%
        \label{fig:host_config}%
        \includegraphics[width=0.8\columnwidth]{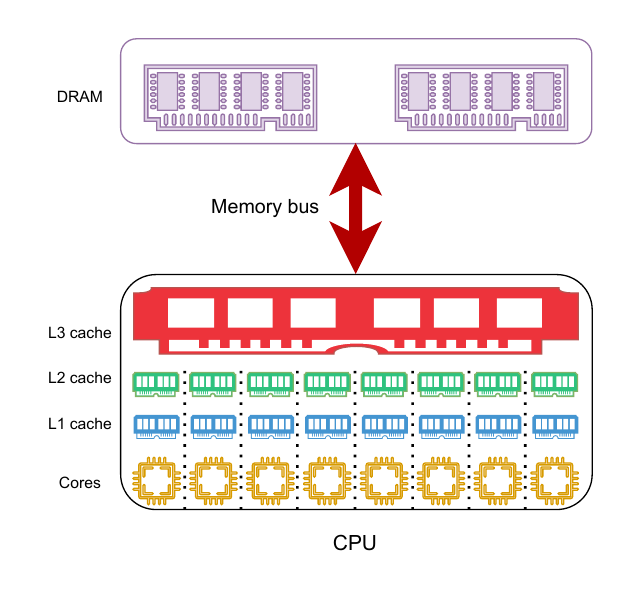}%
    }
    \par\vspace{1em} 
    \subfloat[PIM configuration]{%
        \label{fig:pim_config}%
        \includegraphics[width=0.8\columnwidth]{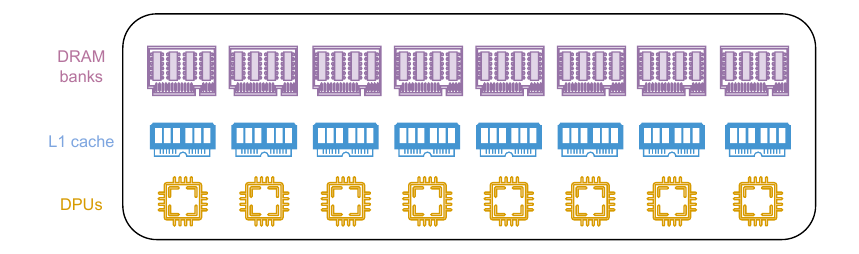}%
    }
    \caption{The memory hierarchy configurations simulated by DAMOV-SIM.}
    \label{fig:damov_configs}
\end{figure}
To simulate how different cache hierarchy configurations affect the performance of a given function, we test two different cache hierarchy configurations as shown in Figure~\ref{fig:damov_configs}: a ``host'' configuration where each core has private L1 and L2 caches, but shares an L3 cache across all cores, and a ``PIM'' configuration where each core only has a private L1 cache. This allows us to observe cases where the given workload does not efficiently utilize higher levels of cache memory. However, the ``PIM'' configuration does not account for possible limitations of real PIM systems, such as a limited instruction set, lower clock speed, and limited communication between DRAM processing units (DPUs) in some processing-near-memory implementations. The impact of these limitations should be investigated separately for a specific accelerator design.

\section{Results}
    \subsection{Whole genome alignment}
We analyzed \verb|minimap2|, \verb|mashmap|, and \verb|wfmash|'s performance while mapping CHM13 and HG002 reads to the hg38 reference. We observe the end-to-end performance of each tool using 8, 16, 32, and 64 threads. Detailed hardware and software specifications of the test system are listed in the Supplementary Material.

\begin{table*}[hbp]
    \centering
    \caption{Number of total elapsed clockticks and CPU wall clock time for each whole genome aligner program.}
    \label{tab:wga_times}

    \subfloat[8 threads]{%
    \label{tab:wga_times_8}
    \resizebox{0.45\linewidth}{!}{
    \begin{tabular}{l|r|r}
         \toprule
         \textbf{Program name} & \textbf{hg38 vs. CHM13} & \textbf{hg38 vs. hg002} \\
         \midrule
         \texttt{minimap2} & $2.37\times10^{13}$ (5008.33s) & $1.86\times10^{13}$  (3828.65s)  \\
         \texttt{mashmap} & $9.11\times10^{13}$ (6637.91s) & $8.74\times10^{13}$  (6416.57s)  \\
         \texttt{wfmash} & $4.80\times10^{13}$ (2480.27s) & $6\times10^{13}$ (3116.45s) \\
         \bottomrule
    \end{tabular}}
    }
    \hfill
    \subfloat[64 threads]{%
    \label{tab:wga_times_64}
    \resizebox{.45\linewidth}{!}{%
    \begin{tabular}{l|r|r}
         \toprule
         \textbf{Program name} & \textbf{hg38 vs. CHM13} & \textbf{hg38 vs. hg002} \\
         \midrule
         \texttt{minimap2} & $2.56\times10^{13}$ (5069.33s) & $2.07\times10^{13}$  (3880.22s)  \\
         \texttt{mashmap} & $9.26\times10^{13}$ (3006.44s) & $8.49\times10^{13}$  (2990.88s)  \\
         \texttt{wfmash} & $8.9\times10^{13}$ (1004.52s) & $1.2\times10^{14}$ (1235.45s)\\
         \bottomrule
    \end{tabular}}
    }
\end{table*}
As shown in Table~\ref{tab:wga_times}, the three whole-genome aligners exhibit varying degrees of parallelism at the same thread counts, leading to discrepancies in wall-clock times. The overall microarchitecture utilization of each tool is available in Table~\ref{tab:wga_total_uarch}. We give more details on each program's performance in its respective subsection. 

We observe that while \verb|minimap2| executes the fewest total instructions, it does not fully use all the hardware threads available to it, regardless of the number of cores in a given machine. This lack of parallelism causes \verb|minimap2| to run slower than other whole-genome aligners on systems with many cores.

\begin{table}[htbp]
    \centering
    \caption{Microarchitecture utilization metrics obtained from each whole genome aligner, using the hg38 vs. CHM13 dataset with 64 threads.}
    \label{tab:wga_total_uarch}
    \resizebox{\linewidth}{!}{%
    \begin{tabular}{l||r|r|r|r|r}
         \toprule
         \textbf{Program name} & \textbf{Retiring} & \textbf{Core Bound} & \textbf{Memory Bound} & \textbf{Front-end Bound} & \textbf{Bad Speculation} \\
         \midrule
         \texttt{minimap2} &  48.9\% & 16.9\% & 19.1\% & 5.4\% & 10.7\% \\ 
         \texttt{mashmap} &  23.3\% & 5.8\% & 16.1\% & 24.5\% & 30.2\% \\ 
         \texttt{wfmash} &  32.6\% & 12\% & 53.4\% & 5.5\% & 0\% \\ 
         \bottomrule
    \end{tabular}}
\end{table}

\subsubsection{minimap2}
\begin{figure}[ht]
    \centering
    \includegraphics[width=\linewidth]{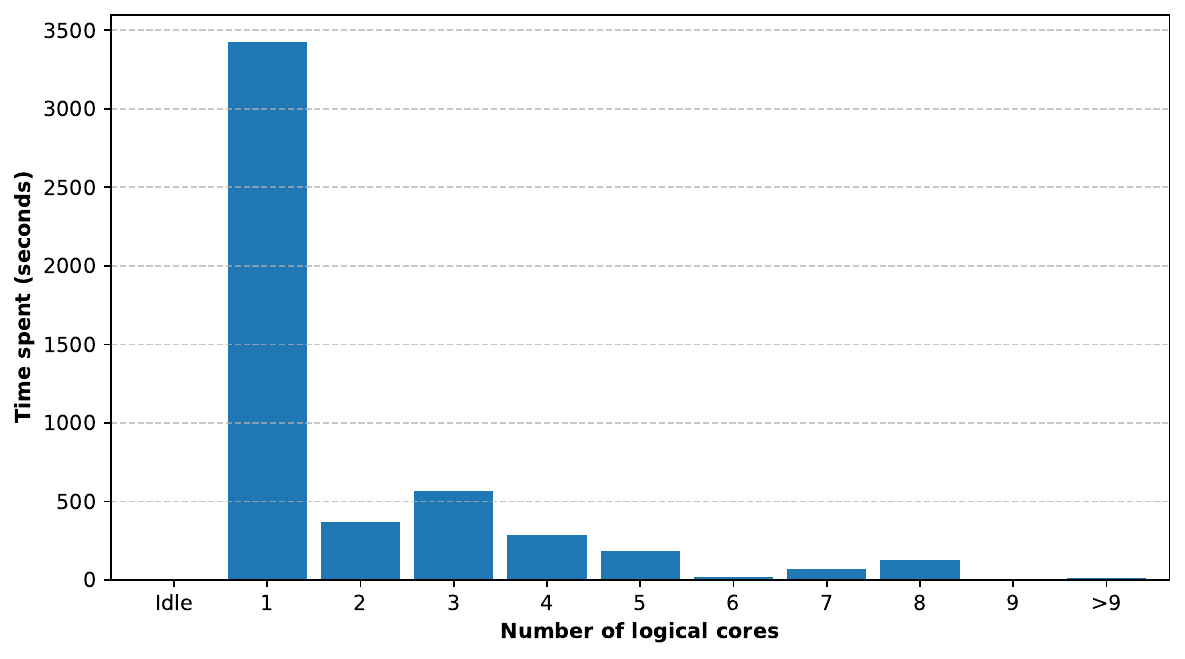}
    \caption{Logical core usage histogram for \texttt{minimap2} on hg38 vs. CHM13 when the program is instructed to use 96 threads.}
    \label{fig:minimap2_wga_hist}
\end{figure}

The most significant performance issue we observe with \texttt{minimap2} in whole genome alignment scenarios is a lack of parallelism, which prevents the program from scaling with the available CPU count, as is apparent in the core usage histogram in Figure~\ref{fig:minimap2_wga_hist}, where it spends 67\% of the CPU time utilizing only a single logical core.

Despite this lack of parallelism, \verb|minimap2| appears to be less bound by microarchitectural inefficiencies than the other two programs, with approximately 49\% of pipeline slots used to successfully retire instructions. The function \verb|ksw_extd2_sse41|, which \verb|minimap2| uses for banded Smith-Waterman alignment, takes 26.4\% of total CPU time and uses 82.1\% of pipeline slots for retiring instructions. As the name suggests, this function also uses the SSE4.1 SIMD extension for instruction-level parallelism. Since the scope of this work is to analyze how these tools perform in a server environment, we did not run tests with fewer than 8 threads. Therefore, it is possible that \verb|minimap2| is more suited for whole genome alignment on environments with few threads available.

\subsubsection{mashmap}
\begin{table*}[!htbp]
    {\centering    
    \caption{Microarchitecture utilization metrics of the top five functions in \texttt{mashmap} (hg38 vs. CHM13 dataset, 64 threads).}
    \label{tab:mashmap_uarch}
    \resizebox{\textwidth}{!}{
    \begin{tabular}{l||r|r|r|r|r|r|r}
         \toprule
         \textbf{Function name} & \textbf{CPU\%} & \textbf{CPI} & \textbf{Retiring} & \textbf{FE} & \textbf{BS} & \textbf{MEM} & \textbf{CORE} \\
         \midrule
         \texttt{std::map<uint32\_t, SlideMapper>::operator[]} & 30.9\% & 1.57 & 13.4\% & 33.1\% & 36\% & 14.8\% & 2.7\% \\ 
         \texttt{SlideMapper::insert\_ref} & 26.8\% & 1.35 & 16.2\% & 30.2\% & 40.8\% & 10.6\% & 2.2\% \\ 
         \texttt{Map::computeL2MappedRegions} & 14.2\% & 1.44 & 15\% & 24.7\% & 35.1\% & 21.8\% & 3.4\% \\ 
         \texttt{std::\_Rb\_tree::erase} & 6.6\% & 1.58 & 12.7\% & 31.1\% & 41\% & 11.8\% & 3.4\% \\ 
         \texttt{Map::mergeMappings} & 3.3\% & 1.02 & 13.5\% & 4.9\% & 10.7\% & 17.1\% & 53.8\% \\ 
         \bottomrule
    \end{tabular}
    }}

    {
    \vspace{.5em}
    \footnotesize  CPU\%: Percentage of total time spent executing the given function, CPI: Clock cycles per executed instruction, FE: Front-end bound, BS: Bad speculation, MEM: Memory-bound, CORE: Core bound.}
\end{table*}
We observe that \verb|mashmap| can meaningfully use up to 24 hardware threads on our server, which explains why it takes less wall-clock time than \verb|minimap2| as the thread count increases. We observe an abnormally high percentage of CPU pipeline stalls due to bad speculation and frontend-bound latency, suggesting unpredictable branching patterns in the \verb|mashmap| codebase. As seen in Table~\ref{tab:mashmap_uarch}, this issue affects all functions that consume significant CPU time, thereby negatively affecting the cycles-per-instruction (CPI) metric. We do not perform any function-level data movement analysis on \verb|mashmap| because none of the functions that take more than 3\% of CPU time are noticeably memory-bound, possibly due to the branch misprediction issue.

\subsubsection{wfmash}

During our testing, we found \verb|wfmash| to be consistently the fastest whole genome alignment program among the three we tested. However, it is also very memory-bound, with 53.4\% of pipeline slots wasted. Examining the memory-bound functions, we observe that an I/O function (\verb|progress_meter::increment|) is extremely memory-bound (97.9\%) and has an average CPI of 82.9. We would like to note that at the time of writing, there is no method (documented or otherwise) to disable the progress meter or otherwise alter the verbosity of the program's output without modifying the source code. The second function of interest is \verb|wavefront_compute_affine2p_idm|, which is a wavefront alignment computation function from WFA2-lib~\cite{wfalib_2021, wfalib_2023}.

\subsection{Long read alignment}
As with whole-genome alignment, we observe that the input size does not meaningfully affect the program's overall performance. We provide results for the VTune analyses using the full SRR1304331 dataset, and for the DAMOV-SIM analyses, we use a randomly sampled subset of 10,000 reads. We shrink the dataset for DAMOV-SIM because the simulation-based approach is orders of magnitude slower than the VTune event counter approach.

\begin{table*}[!htbp]
    \centering
    \caption{Number of total elapsed clockticks and CPU wall clock time for each long read aligner program.}
    \label{tab:long_read_times}
    \resizebox{\linewidth}{!}{
    \begin{tabular}{l|r|r|r|r}
         \toprule
         \textbf{Program name} & \textbf{8 threads} & \textbf{16 threads} & \textbf{32 threads} & \textbf{64 threads} \\
         \midrule
         \texttt{ngmlr} & $4.36\times10^{13}$ (2250.10s) & $4.25\times10^{13}$  (1222.22s) & $4.19\times10^{13}$ (721.673s) & $4.74\times10^{13}$ (483.64s)  \\
         \texttt{minimap2} & $4.33\times10^{12}$ (283.72s) & $4.32\times10^{12}$  (188.77s) & $4.73\times10^{12}$ (151.54s) & $5.27\times10^{12}$ (135.93s)  \\
         \texttt{lra} & $6.53\times10^{12}$ (368.10s) & $6.82\times10^{12}$  (226.57s) & $7.08\times10^{12}$ (180.33s) & $7.28\times10^{12}$ (186.13s)  \\
         \bottomrule
    \end{tabular}
    }
\end{table*}

As seen in Table~\ref{tab:long_read_times}, unlike in the whole genome alignment scenario, all three programs we tested can benefit from increasing the thread count (except \texttt{LRA}, which could not parallelize beyond 24 threads, as reported by VTune). \texttt{NGMLR} was an order of magnitude slower compared to the other tools when all tools were constrained to 8 threads. However, the gap shrank to \texttt{NGMLR} being approximately $4\times$ slower than the competition at a thread count of 64.

\subsubsection{NGMLR}
\texttt{NGMLR} was measured to be the slowest of the three long read aligners we have tested. We hypothesize that this is due to the algorithm \texttt{NGMLR} uses being more sensitive compared to the other aligners. The program executed $1.24\times10^{14}$ instructions for the same dataset and thread count (SRR1304331, 64 threads) compared to minimap2's $8\times10^{12}$ instructions.

\begin{table}[hbt]
    {\centering
    \caption{Microarchitecture utilization metrics of the top five functions in \texttt{ngmlr} (SRR1304331, 64 threads). }
    \label{tab:ngmlr_uarch}
    \resizebox{\linewidth}{!}{
    \begin{tabular}{l||r|r|r|r|r|r|r}
         \toprule
         \textbf{Function name} & \textbf{CPU\%} & \textbf{CPI} & \textbf{Retiring} & \textbf{FE} & \textbf{BS} & \textbf{MEM} & \textbf{CORE} \\
         \midrule
         \texttt{AlignmentBuffer::cLIS} & 49.7\% & 0.31 & 86.4\% & 4.6\% & 6.0\% & 0.6\% & 2.4\% \\ 
         \texttt{Convex::fwdFillMatrixSSESimple} & 16.9\% & 0.4 & 71.9\% & 4.8\% & 12\% & 2.7\% & 8.6\% \\ 
         \texttt{Convex::getDirection} & 4.5\% & 0.35 & 77.2\% & 6.1\% & 9.4\% & 1.9\% & 5.4\% \\ 
         \texttt{CS::AddLocationStd} & 4.3\% & 0.97 & 35.8\% & 3.2\% & 10.7\% & 40.1\% & 10.2\% \\ 
         \texttt{RefEntry::getRealLocation} & 2.6\% & 9.62 & 3.3\% & 0.5\% & 0.2\% & 93.8\% & 2.2\% \\ 
         \bottomrule
    \end{tabular}
    }}
    {\vspace{.2em}
    
    \footnotesize  CPU\%: Percentage of total time spent executing the given function, CPI: Clock cycles per instruction, FE: Front-end bound, BS: Bad speculation, MEM: Memory-bound, CORE: Core bound.}
\end{table}

We did not observe an overall inefficiency in microarchitec\-ture utilization, since both the entire program and the functions that took the most time had a high percentage of retiring pipeline slots ($>70\%$, see Table~\ref{tab:ngmlr_uarch}).

Further inspecting the function-by-function breakdown of execution time on Table~\ref{tab:ngmlr_uarch}, we spot \verb|CS::AddLocationStd| and \texttt{RefEntry::getRealLocation} as highly memory-bound functions. The second function, in particular, has an abnormally high CPI and is 93.8\% memory-bound. 

Searching the \texttt{NGMLR} source code for the function \texttt{RefEntry::getRealLocation} reveals that it is a one-liner that adds an offset to a field in a \verb|Location| struct. Based on this information, we suggest that NGMLR use prefetching, either by reorganizing the data layout to be more cache-friendly on the CPU or by using hardware-software co-design to implement a prefetcher for the \verb|Location| data structure to avoid memory latency. Another hardware-based method for latency reduction would be to use a processing-in-memory framework, such as Ambit~\cite{ambit_2017} or ComputeDRAM~\cite{computedram_2019} (which runs on off-the-shelf DRAM), to perform this simple calculation on the DRAM chip itself rather than moving the data between the CPU and system memory.

\subsubsection{minimap2}

\texttt{minimap2} was the fastest out of the three long read aligners we tested. Unlike whole-genome alignment tests, the program did not suffer as much from poor parallelization, and its retiring pipeline slot rate was similar to that in the whole-genome alignment scenario (49.3\%).

\begin{table}[hbt]
    {\centering
    \caption{Microarchitecture utilization metrics of the top five functions in \texttt{minimap2} (SRR1304331, 64 threads). }
    \label{tab:minimap2_uarch}
    \resizebox{\linewidth}{!}{
    \begin{tabular}{l||r|r|r|r|r|r|r}
         \toprule
         \textbf{Function name} & \textbf{CPU\%} & \textbf{CPI} & \textbf{Retiring} & \textbf{FE} & \textbf{BS} & \textbf{MEM} & \textbf{CORE} \\
         \midrule
         \texttt{ksw\_extd2\_sse41} & 70.9\% & 0.52 & 58.7\% & 2.4\% & 1.5\% & 13.7\% & 23.7\% \\ 
         \texttt{rs\_sort\_128x} & 5.5\% & 0.64 & 43.2\% & 4.6\% & 13.7\% & 22.3\% & 16.2\% \\ 
         \texttt{mm\_sketch} & 3.8\% & 0.5 & 48.3\% & 6.4\% & 31.7\% & 2.9\% & 10.7\% \\ 
         \texttt{mm\_idx\_get} & 3.7\% & 17.6 & 1.8\% & 0.5\% & 1.3\% & 91.3\% & 5.1\% \\ 
         \texttt{collect\_seed\_hits} & 2.6\% & 2.39 & 11.1\% & 6.5\% & 9\% & 55.6\% & 17.1\% \\ 
         \bottomrule
    \end{tabular}
    }}    
    {\vspace{.2em}
    
    \footnotesize  CPU\%: Percentage of total time spent executing the given function, CPI: Clock cycles per instruction, FE: Front-end bound, BS: Bad speculation, MEM: Memory-bound, CORE: Core bound.}
\end{table}

Inspecting Table~\ref{tab:minimap2_uarch}, we see that \verb|ksw_extd2_sse41| still takes most of the execution time and does not exhibit any apparent microarchitectural inefficiencies. The functions \verb|mm_idx_get| and \verb|collect_seed_hits| both suffer from high CPI due to memory-bound latency, warranting further analysis using DAMOV-SIM.

\begin{table}[ht]
    \centering
    \caption{Cache statistics obtained via DAMOV-SIM for the \texttt{mm\_idx\_get} function of \texttt{minimap2}}
    \label{tab:minimap2_damov_idx}
    \subfloat[Host configuration]{
    \label{tab:minimap2_damov_idx_host}
    \begin{tabular}{r||r|r|r|r|r}
        \toprule
         \textbf{Threads} & \textbf{IPC} & \multicolumn{3}{c|}{\textbf{Miss Rates}} & \textbf{LFMR} \\
         & & \textbf{L1} & \textbf{L2} & \textbf{L3} & \\
         \midrule
         1 & 1.39 & 24.1\% & 61.9\% & 14.2\% & 0.09 \\
         4 & 5.16 & 24.1\% & 62.2\% & 25.5\% & 0.174 \\
         16 & 16.6 & 24.1\% & 62.5\% & 56.4\% & 0.386 \\
         64 & 30.5 & 24.4\% & 72.5\% & 96.1\% & 0.741 \\
         128 & 31.2 & 24.7\% & 84.6\% & 96.0\% & 0.838 \\
         256 & 28.2 & 26.3\% & 91.9\% & 95.0\% & 0.887 \\
         \bottomrule
    \end{tabular} 
    }
    
    \subfloat[PIM configuration]{
    \label{tab:minimap2_damov_idx_pim}
    \begin{tabular}{r||r|r}
        \toprule
         \textbf{Threads} & \textbf{IPC} & \textbf{L1 Miss Rate}  \\
         \midrule
         1 & 1.21 & 24.1\% \\ 
         4 & 4.54 & 24.1\% \\
         16 & 15.1 & 24.1\% \\
         64 & 29.5 & 24.1\% \\
         128 & 33.8 & 24.2\% \\
         256 & 34.1 & 25.4\% \\
         \bottomrule
    \end{tabular}
    }
\end{table}

Upon further analysis using DAMOV-SIM, we observe that both functions exhibit increasing LFMR (last-to-first cache miss ratio) as thread size increases under the ``host'' configuration (see Table~\ref{tab:minimap2_damov_idx_host}). This alludes to the last level of the cache hierarchy becoming less useful as the core count increases, where checking and repopulating L3 cache entries becomes an architectural burden on the simulated system. Specifically for \verb|mm_idx_get|, these cache misses cause the instructions per cycle (IPC) metric to stall at around 30 for thread counts $\ge64$. (DAMOV-SIM's IPC calculation method is different from VTune's CPI calculation, which means that the ideal IPC for DAMOV-SIM is equal to the core count.) We rerun the DAMOV-SIM experiments using the simpler ``PIM'' cache configuration and observe a slight improvement in the IPC metric, as seen in Table~\ref{tab:minimap2_damov_idx_pim}. This shows that a multi-level, general-purpose cache configuration causes latency issues as thread counts increase, rather than closing the speed gap between the CPU and RAM.
\subsubsection{LRA}

While running LRA for long read alignment, the function \verb|CompareLists<GenomeTuple,uint64_t>| took the most time to execute, taking 27.1\% of the execution time. This function had an abnormally high CPI of 11.2 and was 81\% memory-bound. LRA also suffers the most from microarchitectural inefficiencies across the entire program, with the lowest retiring-pipeline-slot rate among the long read aligners at 27\%. 

\begin{table}[ht]
    \centering
    \caption{Cache statistics obtained via DAMOV-SIM for the \texttt{CompareLists} function of \texttt{lra}}
    \label{tab:lra_damov_host}
    \begin{tabular}{r||r|r|r|r|r}
        \toprule
         \textbf{Threads} & \textbf{IPC} & \multicolumn{3}{c|}{\textbf{Miss Rates}} & \textbf{LFMR} \\
         & & \textbf{L1} & \textbf{L2} & \textbf{L3} & \\
         \midrule
         1 & 0.567 & 22.7\% & 49.4\% & 51.5\% & 0.373 \\
         4 & 1.99 & 18.5\% & 46.1\% & 63.3\% & 0.420 \\
         16 & 2.32 & 18.4\% & 45.2\% & 65.2\% & 0.430 \\
         64 & 2.24 & 18.2\% & 46.5\% & 73.2\% & 0.491 \\
         128 & 2.23 & 18.7\% & 43.9\% & 73.4\% & 0.478 \\
         \bottomrule
    \end{tabular}
\end{table}

We further analyzed \verb|CompareLists| using DAMOV-SIM, revealing a less pronounced variant of the pattern observed in minimap2's memory-bottlenecked functions as seen on Table~\ref{tab:lra_damov_host}.  The L1 cache miss rate stabilizes at approximately 18\% for multi-core experiments on the ``host'' configuration, while the L3 cache miss rate (shared across cores) continues to climb as the core count increases. Meanwhile, the L2 miss rate decreases slightly, suggesting an issue with the shared nature of the last-level cache. These problems cause the IPC value to stall between 1.98 and 2.32 in the multi-core experiments, consistent with the limited parallelism we observed in the VTune results.

\subsection{Short read alignment}
We test four short-read alignment programs: \texttt{StrobeAlign}, \texttt{bowtie2}, \texttt{BWA-MEM}, and \texttt{minibwa}. 
We used several reads from the CHM1 dataset (Eichler-CHM1htert.1\_500 library25) with SRA IDs SRR1514950-1514952 to analyze all three aligners. The read sets are pair-ended, each averaging 43G bases and providing a fold coverage of $41\times$. We let all three programs construct indexes of the reference prior to the performance measurements.

\begin{table*}[hbp]
    \centering
    \caption{Number of total elapsed clockticks and CPU wall clock time for each short-read aligner program (average of SRR1514\{950,951,952\}.}
    \label{tab:short_read_times}
    \resizebox{\linewidth}{!}{
    \begin{tabular}{l|r|r|r|r}
         \toprule
         \textbf{Program name} & \textbf{8 threads} & \textbf{16 threads} & \textbf{32 threads} & \textbf{64 threads} \\
         \midrule
         \texttt{strobealign} & $7.83\times10^{13}$ ($3.97\times10^{3}$s) & $7.99\times10^{12}$  ($2.10\times10^{3}$s) & $8.41\times10^{13}$ ($2.04\times10^{3}$s) & $4.43\times10^{13}$ ($1.44\times10^{3}$s)  \\
         \texttt{bowtie2} & $3.41\times10^{14}$ ($1.91\times10^{4}$s) & $3.64\times10^{14}$  ($1.01\times10^{4}$s) & $3.70\times10^{14}$ ($5.16\times10^{3}$s) & $4.57\times10^{14}$ ($3.02\times10^{3}$s)  \\
         \texttt{bwa-mem} & $3.81\times10^{14}$ ($2.24\times10^{4}$s) & $3.98\times10^{14}$  ($1.17\times10^{4}$s) & $4.32\times10^{14}$ ($6.12\times10^{3}$s) & $4.43\times10^{14}$ ($4.59\times10^{3}$s)  \\
         \texttt{minibwa} & $8.28\times10^{13}$ ($4.18\times10^{3}$) & $8.14\times10^{13}$ ($2.42\times10^{3}$s) & $1\times10^{14}$ ($1.58\times10^{3}$s) & $7.98\times10^{13}$ ($2.23\times10^{3}$s) \\
         \bottomrule
    \end{tabular}
    }
\end{table*}

\begin{table}[htbp]
    \centering
    \caption{Microarchitecture utilization metrics obtained from each short-read aligner (SRR1514950, 64 threads).}
    \label{tab:short_total_uarch}
    \resizebox{\linewidth}{!}{%
    \begin{tabular}{l||r|r|r|r|r}
         \toprule
         \textbf{Program name} & \textbf{Retiring} & \textbf{Core Bound} & \textbf{Memory Bound} & \textbf{Front-end Bound} & \textbf{Bad Speculation} \\
         \midrule
         \texttt{strobealign} & 42\% & 10.6\% & 23.3\% & 10.2\% & 13.9\% \\
         \texttt{bowtie2} &  34.8\% & 16.4\% & 31.8\% & 10.2\% & 6.8\% \\ 
         \texttt{bwa-mem} &  35.1\% & 8.5\% & 45.6\% & 3.9\% & 6.8\% \\
         \texttt{minibwa} & 41.9\% & 19.0\% & 22.8\% & 7.0\% & 9.3\% \\
         \bottomrule
    \end{tabular}
    }
\end{table}

As seen in Table~\ref{tab:short_read_times}, \texttt{StrobeAlign} is the fastest among the short read mappers we have tested, followed by \texttt{minibwa}, except for the runs using 32 threads, where \texttt{minibwa} is faster than \texttt{StrobeAlign}. We further observe that, based on the microarchitecture utilization metrics shown in Table~\ref{tab:short_total_uarch}, \texttt{StrobeAlign} has the highest percentage of retiring pipeline slots, closely followed by \texttt{minibwa}. \texttt{Bowtie2} and \texttt{BWA-MEM} have similar retirement rates; however, \texttt{BWA-MEM} is overall more memory-bound, with a 45.6\% memory-boundness rate. 

\subsubsection{StrobeAlign}

We observe diminishing returns in \texttt{StrobeAlign}'s performan\-ce at thread counts above 16. 
On our test machine with four 14-core processors with hyperthreading, using more than 28 threads requires spreading them across multiple processors. This might incur additional memory-bound delays during execution due to non-uniform memory access (NUMA), where requested data may reside on a different core and require additional data movement.

\begin{table}[hbt]
    {\centering
    \caption{Microarchitecture utilization metrics of the top five functions in StrobeAlign (SRR1514950, 64 threads).}
    \label{tab:strobealign_uarch}
    \resizebox{\linewidth}{!}{
    \begin{tabular}{l||r|r|r|r|r|r|r}
         \toprule
         \textbf{Function name} & \textbf{CPU\%} & \textbf{CPI} & \textbf{Retiring} & \textbf{FE} & \textbf{BS} & \textbf{MEM} & \textbf{CORE} \\
         \midrule
         \texttt{sw\_sse2\_byte} & 26.1\% & 0.34 & 71.7\% & 8.9\% & 11.1\% & 1.0\% & 7.2\% \\ 
         \texttt{StrobemerIndex::get\_masked\_from} & 15.9\%  & 5.303 & 4.3\% & 2.0\% & 9.9\% & 78.7\% & 5.0\% \\ 
         \texttt{ssw\_align} & 6.1\% & 0.394 & 69.3\% & 7.6\% & 12.6\% & 1.6\% & 8.9\% \\ 
         \texttt{memmem::searcher} & 5.8\% & 1.082 & 24.1\% & 9.3\% & 45.5\% & 12.0\% & 9.1\% \\ 
         \texttt{get\_nams\_by\_chaining} & 5.5\% & 0.577 & 44.1\% & 11.3\% & 18.9\% & 12.6\% & 13.1\% \\ 
         \bottomrule
    \end{tabular}
    }}
    {\vspace{.2em}
    
    \footnotesize  CPU\%: Percentage of total time spent executing the given function, CPI: Clock cycles per instruction, FE: Front-end bound, BS: Bad speculation, MEM: Memory-bound, CORE: Core bound.}
\end{table}

Investigating the profiler data for \texttt{StrobeAlign} reveals that the experiments using more than 8 CPU threads report a high overall percentage of remote memory accesses, which contributes to the high memory boundness percentage of the \texttt{StrobemerIndex::get\_masked\_from} function, as seen in Table~\ref{tab:strobealign_uarch}. At low thread counts, pinning \texttt{StrobeAlign}'s CPU and memory usage to a single NUMA node reduces overall memory boundness but does not yield a significant decrease in execution time.

\begin{table}[hbt]
    {\centering
    \caption{Effect of pinning StrobeAlign to a single NUMA node in low thread counts (SRR1514950). }
    \label{tab:strobealign_numa}
    \resizebox{\linewidth}{!}{
    \begin{tabular}{r|c||r|r|r}
         \toprule
         \textbf{Thread count} & \textbf{Pinned?} & \textbf{Remote} & \textbf{MEM} & \textbf{Time (s)} \\
         \midrule
         8 & No & 54.3\% & 18.1\% & 3938 \\
         8 & Yes & 1.8\% & 14.1\% & 3876\\
         14 & No & 70.9\% & 21.6\% & 2623 \\
         14 & Yes & 3.6\% & 14.5\% & 2504\\
         \bottomrule
    \end{tabular}
    }}
    {\vspace{.2em}
    
    \footnotesize  Remote: Remote memory access percentage, MEM: Overall memory boundness, Time: Wall clock time}
\end{table}

\subsubsection{Bowtie2}
We observe that the functions \texttt{SideLocus::initFromRow} and \texttt{AlignmentCache::addOnTheFlyImpl} are memory-bound in Bowtie2, as seen in Table~\ref{tab:bowtie2_uarch}. Hotspot analysis on \texttt{SideLocus::initFromRow} reveals that the memory boundness of this function is caused by calls to \texttt{\_\_builtin\_prefetch}. Disabling prefetching inside this function reduces the function's memory boundness but does not meaningfully affect overall memory boundness or execution time.

\begin{table}[hbt]
    {\centering
    \caption{Microarchitecture utilization metrics of the top five functions in Bowtie2 (SRR1514950, 64 threads).}
    \label{tab:bowtie2_uarch}
    \resizebox{\linewidth}{!}{
    \begin{tabular}{l||r|r|r|r|r|r|r}
         \toprule
         \textbf{Function name} & \textbf{CPU\%} & \textbf{CPI} & \textbf{Retiring} & \textbf{FE} & \textbf{BS} & \textbf{MEM} & \textbf{CORE} \\
         \midrule
         \texttt{SwAligner::alignNucleotidesEnd2EndSseU8} & 11.3\% & 0.466 & 63.4\% & 8.6\% & 7.8\% & 8.5\% & 11.6\% \\ 
         \texttt{SideLocus::initFromRow} & 9.9\% & 4.67 & 6.6\% & 4.3\% & 11.0\% & 54.4\% & 23.8\% \\ 
         \texttt{AlignmentCache::addOnTheFlyImpl} & 5.9\% & 1.078 & 17.4\% & 0.5\% & 2.7\% & 42.7\% & 36.6\% \\ 
         \texttt{\_\_mm\_max\_epu8} & 4.4\% & 0.376 & 74.3\% & 8.0\% & 4.3\% & 5.6\% & 7.8\% \\ 
         \texttt{\_\_mm\_subs\_epu8} & 4.1\% & 0.379 & 73.2\% & 8.2\% & 0.5\% & 7.7\% & 10.4\% \\ 
         \bottomrule
    \end{tabular}
    }}
    {\vspace{.2em}
    
    \footnotesize  CPU\%: Percentage of total time spent executing the given function, CPI: Clock cycles per instruction, FE: Front-end bound, BS: Bad speculation, MEM: Memory-bound, CORE: Core bound.}
\end{table}

\subsubsection{BWA-MEM}

\begin{table}[hbt]
    {\centering
    \caption{Microarchitecture utilization metrics of the top five functions in \texttt{BWA-MEM} (SRR1514950, 64 threads).}
    \label{tab:bwamem_uarch}
    \resizebox{\linewidth}{!}{
    \begin{tabular}{l||r|r|r|r|r|r|r}
         \toprule
         \textbf{Function name} & \textbf{CPU\%} & \textbf{CPI} & \textbf{Retiring} & \textbf{FE} & \textbf{BS} & \textbf{MEM} & \textbf{CORE} \\
         \midrule
         \texttt{bwt\_sa} & 23.6\% & 4.36 & 7.2\% & 1.4\% & 3.4\% & 83.6\% & 4.4\% \\ 
         \texttt{bwt\_occ4} & 15.9\% & 1.63 & 17.4\% & 0.6\% & 3.4\% & 71.9\% & 6.7\% \\ 
         \texttt{ksw\_extend2} & 15.5\% & 0.44 & 76.4\% & 2.7\% & 10.3\% & 2.3\% & 8.3\% \\ 
         \texttt{ksw\_u8} & 10.8\% & 0.34 & 75.8\% & 4.9\% & 4.9\% & 1.6\% & 12.7\% \\ 
         \texttt{bwt\_2occ4} & 8.7\% & 1.751 & 14.4\% & 1.0\% & 6.9\% & 73.6\% & 4.0\% \\ 
         \bottomrule
    \end{tabular}
    }}    
    {\vspace{.2em}
    
    \footnotesize  CPU\%: Percentage of total time spent executing the given function, CPI: Clock cycles per instruction, FE: Front-end bound, BS: Bad speculation, MEM: Memory-bound, CORE: Core bound.}
\end{table}

We identify two target functions in \texttt{BWA-MEM} for in-depth analysis: \verb|bwt_sa|, used for suffix array lookup; and \verb|ksw_extend2|, which is used for the banded Smith-Waterman alignment algorithm. We switch to using reads simulated from \textit{C. elegans} for the function-level performance characterization, since the simulator-based workflow is orders of magnitude slower than VTune. We simulate reads using \verb|art_illumina|~\cite{art_2011}, generating pair-end reads that are 150 bp on average with $2\times$ coverage. We use \verb|GCF_000002985.6_WBcel235| as our reference.

\begin{table}[htbp]
    \centering
    \caption{Cache statistics obtained via DAMOV-SIM for the \texttt{bwt\_sa} function of \texttt{BWA-MEM}, using the ``host'' cache layout}
    \label{tab:bwa_damov_bwt_sa}
    \subfloat[Host configuration]{
    \label{tab:bwa_damov_bwt_sa_host}
    \begin{tabular}{r||r|r|r|r|r}
        \toprule
         \textbf{Threads} & \textbf{IPC} & \multicolumn{3}{c|}{\textbf{Miss Rates}} & \textbf{LFMR} \\
         & & \textbf{L1} & \textbf{L2} & \textbf{L3} & \\
         \midrule
         1 & 1.63 & 10.1\% & 54.3\% & 70.2\% & 0.388 \\
         4 & 6.48 & 10.2\% & 55.4\% & 68.5\% & 0.392 \\
         16 & 24.7 & 10.7\% & 54.3\% & 70.0\% & 0.393 \\
         64 & 43.9 & 10.3\% & 56.5\% & 72.9\% & 0.428 \\
         128 & 55.1 & 10.5\% & 60.4\% & 75.1\% & 0.474 \\
         \bottomrule
    \end{tabular}
    
    }
    
    \subfloat[PIM configuration]{
    \label{tab:bwa_damov_bwt_sa_pim}
    \begin{tabular}{r||r|r}
        \toprule
         \textbf{Threads} & \textbf{IPC} & \textbf{L1 Miss Rate}  \\
         \midrule
         1 & 1.66 & 10.4\% \\ 
         4 & 6.40 & 10.5\% \\
         16 & 24.8 & 10.9\% \\
         64 & 80.0 & 10.6\% \\
         128 & 125 & 10.7\% \\
         \bottomrule
    \end{tabular}
    }
\end{table}

\begin{figure}
    \centering
    \includegraphics[width=\linewidth]{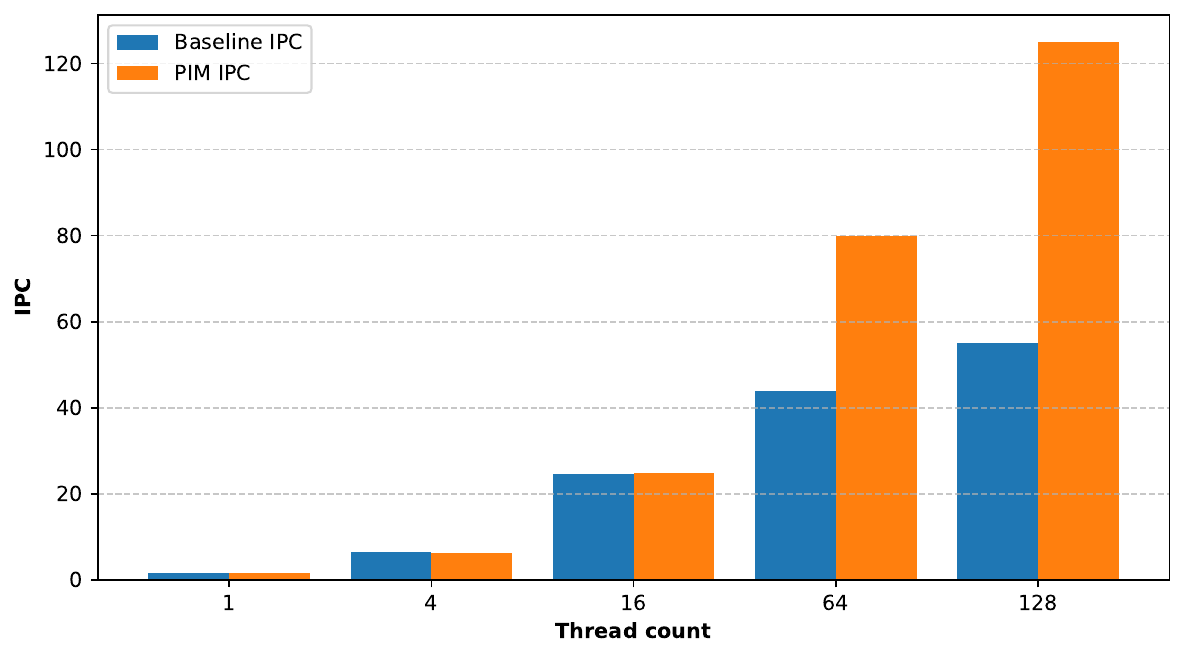}
    \caption{IPC comparison for \texttt{bwt\_sa}, under the ``host'' and ``PIM'' configurations of DAMOV-SIM}
    \label{fig:bwa_ipcs}
\end{figure}

As seen on Table~\ref{tab:bwa_damov_bwt_sa_host}, \verb|bwt_sa| also suffers from last-level-cache induced latency and its IPC starts to plateau as we increase the thread count. Upon switching to the simpler  ``PIM'' cache layout (see Table~\ref{tab:bwa_damov_bwt_sa_pim} and Figure~\ref{fig:bwa_ipcs}), we see a remarkable increase in IPC for high thread counts, which also points to L3 cache contention causing latency problems for \texttt{BWA-MEM}. These findings are consistent with previous literature. Daib et al.~\cite{realpim_2023} observed that a system with lower thread count but a bigger L3 cache was more performant compared to a similar system with more CPU cores but a smaller L3 cache.

\subsubsection{minibwa}

\begin{table}[hbt]
    {\centering
    \caption{Microarchitecture utilization metrics of the top five functions in \texttt{minibwa} (SRR1514950, 64 threads).}
    \label{tab:minibwa_uarch}
    \resizebox{\linewidth}{!}{
    \begin{tabular}{l||r|r|r|r|r|r|r}
         \toprule
         \textbf{Function name} & \textbf{CPU\%} & \textbf{CPI} & \textbf{Retiring} & \textbf{FE} & \textbf{BS} & \textbf{MEM} & \textbf{CORE} \\
         \midrule
         \texttt{ksw\_extd2\_sse} & 32.6\% & 0.50 & 55.3\% & 7.9\% & 9.9\% & 8.0\% & 18.8\% \\ 
         \texttt{mb\_bwt\_sa\_batch} & 10.5\% & 1.478 & 17.7\% & 3.0\% & 8.9\% & 48.4\% & 22.0\% \\ 
         \texttt{ksw\_ll\_u8\_core} & 7.4\% & 0.363 & 53.7\% & 3.2\% & 6.4\% & 4.6\% & 32.0\% \\ 
         \texttt{mb\_bwt\_rank1a} & 7.3\% & 0.653 & 41.3\% & 2.4\% & 2.5\% & 32.9\% & 20.9\% \\ 
         \texttt{mb\_bwt\_smem\_batch} & 5.6\% & 0.944 & 28.9\% & 5.2\% & 27.4\% & 13.7\% & 24.8\% \\ 
         \bottomrule
    \end{tabular}
    }}
    {\vspace{.2em}
    
    \footnotesize  CPU\%: Percentage of total time spent executing the given function, CPI: Clock cycles per instruction, FE: Front-end bound, BS: Bad speculation, MEM: Memory-bound, CORE: Core bound.}
\end{table}

\texttt{minibwa}~\cite{minibwa} is a short read mapper that improves upon \texttt{BWA-MEM} by combining the seeding step from \texttt{BWA-MEM2} with the seed extension and alignment steps from \texttt{minimap2}. We observe that these design changes translate into improved performance compared to both the original \texttt{BWA-MEM} and \texttt{Bowtie2}. In particular, the batched suffix array lookup function \texttt{mb\_bwt\_sa\_batch} is approximately $0.6\times$ as memory-bound as its counterpart in \texttt{BWA-MEM} (\texttt{bwt\_sa}). This contributes to a reduced overall memory boundness percentage and to less time spent on suffix array lookups.

\section{Conclusions and future work}
    The aim of this study is to analyze the performance and data movement characteristics of DNA read mapping tools as the programs scale to higher thread counts. To achieve this, we profiled nine tools across three read mapping categories and collected their microarchitecture utilization metrics. We further analyzed specific memory-bound functions using a simulation-based workflow to observe their cache utilization patterns in detail.

We observe that microarchitectural inefficiencies, including memory-bound functions, and poor parallelism are the main culprits of poor performance across whole genome, long read, and short read alignment tools. We combine results from both real server-grade hardware and in-depth memory and CPU simulations to reveal that DNA read mapping tools are likely to suffer from memory-related bottlenecks due to L3 cache contention, with \texttt{LRA} and \texttt{BWA-MEM} being the most prominent examples. We hypothesize that such bottlenecks can be alleviated by offloading memory-intensive parts of these programs onto accelerator circuitry closer to system memory, thereby avoiding data movement bottlenecks.

The scope of our experiments was limited to highly-parallelized, server-grade systems. As we noted in Section 3.1.1, tools like \verb|minimap2| could prove to be more efficient for systems with very limited core counts, since the program is highly vectorized and does not utilize many CPU cores during whole genome alignment. We also came across obstacles while performing function-level analysis on \texttt{mashmap} and \texttt{NGMLR} due to some bugs in Zsim. Since improving the DAMOV-SIM simulator was out of the scope of this article, we did not fully investigate the Zsim or Ramulator codebases. Future work could improve upon the DAMOV-SIM workflow by augmenting or replacing the DRAM and CPU simulators with their more up-to-date counterparts.

\section{Conflicts of Interest}
C. Alkan is a co-founder of Lidya Genomics.

\section{Acknowledgments} 
This work was supported by the European Union’s Horizon Programme for Research and Innovation under grant agreement No. 101047160 (Project BioPIM).

\bibliographystyle{unsrt}
\bibliography{calkan,references}

\onecolumn
\newpage
\pagenumbering{Roman}
\setcounter{page}{0}
\setcounter{figure}{0}
\setcounter{table}{0}
\setcounter{section}{0}
\setcounter{subsection}{0}
\renewcommand{\thefigure}{S\arabic{figure}}
\renewcommand{\thetable}{S\arabic{table}}
\renewcommand{\thesubsection}{S\arabic{subsection}}

\titleformat{\subsection}
  {\normalfont\bfseries} 
  {\thesubsection:}               
  {1em}                        
  {}

\section{Supplementary Materials}

\subsection{Test environment}
\subsubsection{Program versions}
\begin{itemize}
    \item \textbf{minimap2:} 2.26-r1175
    \item \textbf{mashmap:} v2.0
    \item \textbf{wfmash:} v0.24.1
    \item \textbf{NGMLR:} 0.2.7
    \item \textbf{LRA:} 1.3.5
    \item \textbf{BWA-MEM:} 0.7.17
    \item \textbf{bowtie2:} 2.5.1
    \item \textbf{StrobeAlign:} v0.17.1
    \item \textbf{minibwa:} 0.3 (r391)
    \item \textbf{Analysis tools:}
    \begin{itemize}
        \item \textbf{VTune:} 2025.0
        \item \textbf{DAMOV-SIM:} Latest release modified for compatibility with later Ubuntu versions. Available at \url{https://github.com/biopim/DAMOV}
    \end{itemize}
\end{itemize}
\subsubsection{Server specifications}
\begin{itemize}
    \item $4\times$Intel(R) Xeon(R) CPU E7-4850 v3 @ 2.20GHz
    \begin{itemize}
        \item Number of cores:	14
        \item Number of hyperthreads: 28
        \item Supported SIMD extensions: \texttt{mmx, sse, sse2, sse3, ssse3, sse4.1, sse4.2, avx, avx2}
    \end{itemize}
    \item 1 TB system memory
    \item Operating system: Ubuntu 22.04
    \begin{itemize}
        \item Kernel version: 5.15.0-131-generic
    \end{itemize}

\end{itemize}
\subsection{Program invocations}
\subsubsection{Whole genome alignment}

\paragraph{minimap2}

\begin{verbatim}
sudo /opt/intel/oneapi/vtune/latest/bin64/vtune -collect memory-access -data-limit=20000 \
/mnt/compgen/inhouse/bin/minimap2 -ax asm5 -t ${SLURM_CPUS_PER_TASK} ${GRCh38} ${QUERY_FILE} \
> ${OUTPUT_DIR}/${TOOL_NAME}_grch38_${QUERY_FILENAME}_t${SLURM_CPUS_PER_TASK}.sam
\end{verbatim}

\paragraph{mashmap}

\begin{verbatim}
sudo /opt/intel/oneapi/vtune/latest/bin64/vtune -collect memory-access -data-limit=20000 \
/mnt/compgen/inhouse/bin/mashmap -t ${SLURM_CPUS_PER_TASK} -r ${GRCh38} -q ${QUERY_FILE} \
-o ${OUTPUT_DIR}/mashmap_grch38_${QUERY_FILENAME}_t${SLURM_CPUS_PER_TASK}.mashmapout \
> ${OUTPUT_DIR}/mashmap_grch38_${QUERY_FILENAME}_t${SLURM_CPUS_PER_TASK}.stdout
\end{verbatim}

\paragraph{wfmash}
\begin{verbatim}
sudo ${VTUNE_CMD} -collect memory-access -result-dir \
${VTUNE_OUT} -data-limit=100000 \
wfmash ${GRCh38} ${QUERY_FILE} \
-p 99 --threads ${SLURM_CPUS_PER_TASK} \
> ${OUTPUT_DIR}/${EXPERIMENT_NAME}.paf
\end{verbatim}

\subsubsection{Long read alignment}

\paragraph{NGMLR}
\begin{verbatim}
zcat /mnt/storage1/projects/human-cell-lines/chm1/pacbio/fastq/SRR1304331.fastq.gz | \
sudo /opt/intel/oneapi/vtune/latest/bin64/vtune -collect \
memory-access -data-limit=20000 ./ngmlr-0.2.7/ngmlr -t 8 -r \
/mnt/compgen/inhouse/share/rg_annot/b38/human_v38.fasta -q \
/dev/stdin -o ./vtune_sam_outputs/ngmlr_fulldataset_t8.sam > \
./biopim/vtune-reports/memory-access/vtune-ngmlr-t8-human-SRR31-memory-access.txt
\end{verbatim}

\paragraph{minimap2}
\begin{verbatim}
zcat /mnt/storage1/projects/human-cell-lines/chm1/pacbio/fastq/SRR1304331.fastq.gz | \ 
sudo /opt/intel/oneapi/vtune/latest/bin64/vtune -collect memory-access \
./minimap2/minimap2 -ax map-pb -t 8 \
/mnt/compgen/inhouse/share/rg_annot/b38/human_v38.fasta /dev/stdin > \
./vtune_sam_outputs/minimap2_fulldataset_t8.sam 
\end{verbatim}

\paragraph{LRA}
\begin{verbatim}
zcat /mnt/storage1/projects/human-cell-lines/chm1/pacbio/fastq/SRR1304331.fastq.gz | \ 
sudo /opt/intel/oneapi/vtune/latest/bin64/vtune -collect memory-access \
/mnt/compgen/inhouse/bin/lra align -CLR \
/mnt/compgen/inhouse/share/rg_annot/b38/human_v38.fasta \
/dev/stdin -t 8 -p s > ./vtune_sam_outputs/lra_fulldataset_t8.sam 
\end{verbatim}

\subsubsection{Short read alignment}

\paragraph{BWA-MEM}
\begin{verbatim}
PIM_ROOT/bwa-0.7.17/bwa mem -t THREADS \
PIM_ROOT/../data/celegans/GCF_000002985.6_WBcel235_genomic.fna.gz \
PIM_ROOT/../data/celegans/paired_dat1.fq.gz \
PIM_ROOT/../data/celegans/paired_dat2.fq.gz -o \
PIM_ROOT/../data/celegans/paired_THREADS.sam
\end{verbatim}

\paragraph{Bowtie2}
\begin{verbatim}
/home/bsahin/DAMOV/workloads/bowtie2-2.5.4/bowtie2 "-x" \ "./data/chm1_reads/CHM1_1.1" "-1" \ 
"./data/chm1_reads/SRR1514950_CHM1_Illumina_WGS_UW__1.fastq.gz" \
"-2" "./data/chm1_reads/SRR1514950_CHM1_Illumina_WGS_UW__2.fastq.gz" \
"-p" "64" "-t" "--met-file" "workloads/bowtie2-2.5.4/chm1_metrics.log" "-S" "/dev/null" 
\end{verbatim}

\paragraph{StrobeAlign}
\begin{verbatim}
sudo /opt/intel/oneapi/vtune/latest/bin64/vtune -collect memory-access -data-limit=20000 \
  ${STROBEALIGN}$ --use-index -t ${SLURM_CPUS_PER_TASK} ${CHM1} ${QUERY1} ${QUERY2} > \
  ${OUTPUT_DIR}/${TOOL_NAME}_chm1_${QUERY_FILENAME}_t${SLURM_CPUS_PER_TASK}.sam
\end{verbatim}

\paragraph{minibwa}
\begin{verbatim}
sudo /opt/intel/oneapi/vtune/latest/bin64/vtune -collect memory-access -data-limit=20000 \
  ${MINIBWA} mem -t ${SLURM_CPUS_PER_TASK} ${CHM1} ${QUERY1} ${QUERY2} -p -o \
  ${OUTPUT_DIR}/${TOOL_NAME}_chm1_${QUERY_FILENAME}_t${SLURM_CPUS_PER_TASK}.sam
\end{verbatim}

\end{document}